\documentclass{article}

\usepackage{graphicx} 
\usepackage[margin=1in]{geometry} 

\usepackage{hyperref}
\usepackage{algorithm}
\usepackage{algpseudocode}
\usepackage{comment}
\usepackage{bbm}
\usepackage[dvipsnames, table]{xcolor}
\usepackage{mathtools} 
\usepackage{natbib}
\usepackage{graphicx} 
\usepackage{listings}
\usepackage{enumerate}
\usepackage{tabularx}
\usepackage{scalerel} 
\usepackage{amsmath}
\usepackage{amssymb}
\usepackage{amsfonts}
\usepackage{booktabs}
\usepackage{multirow}
\usepackage{threeparttable}
\usepackage{subcaption}

\usepackage{float}
\usepackage{caption}
\usepackage{colortbl}
\newcommand{\nothere}[1]{}

\usepackage{orcidlink}

\title{ 
Penalized Likelihood Optimization for Adaptive Neighborhood Clustering in Time-to-Event Data with Group-Level Heterogeneity}

\author{Alessandra Ragni$^{1}$\orcidlink{0000-0002-3647-7340}, Lara Cavinato$^{1}$\orcidlink{0000-0001-7007-0205}, Francesca Ieva$^{1,2}$\orcidlink{0000-0003-0165-1983}
\\
\small{$^1$MOX, Department of Mathematics, Politecnico di Milano, Piazza Leonardo da Vinci 32, 20133, Milano, Italy} \\
\small{$^2$Health Data Science Centre, Human Technopole, Viale Rita Levi–Montalcini 1, Milan 20157, Italy}}
\date{}

\begin{document}

\maketitle

\abstract{
The identification of patient subgroups with comparable event-risk dynamics 
plays a key role in supporting informed decision-making in clinical research.
In such settings, it is important to account for the inherent 
dependence that arises when individuals are nested within higher-level units, such as hospitals.
Existing survival models account for group-level heterogeneity through frailty terms but do not uncover latent patient subgroups, while most clustering methods ignore hierarchical structure and are not estimated jointly with survival outcomes. In this work, we introduce a new framework that simultaneously performs patient clustering and shared-frailty survival modeling through a penalized likelihood approach. The proposed methodology adaptively learns a patient-to-patient similarity matrix via a modified version of spectral clustering, enabling cluster formation directly from estimated risk profiles while accounting for group membership.
A simulation study highlights the proposed model's ability to recover latent clusters and to correctly estimate hazard parameters.
We apply our method to a large cohort of heart-failure patients hospitalized with COVID-19 between 2020 and 2021 in the Lombardy region (Italy), identifying clinically meaningful subgroups characterized by distinct risk profiles and highlighting the role of respiratory comorbidities and hospital-level variability in shaping mortality outcomes. 
This framework provides a flexible and interpretable tool for risk-based patient stratification in hierarchical data settings.
}

\vspace{0.3cm}


\section{Introduction}
\label{sec:introduction}




In real-world time-to-event studies arising from clinical and epidemiological studies, identifying patients with similar risk trajectories enables the formation of subgroups, or clusters, characterized by shared likelihoods of experiencing key outcomes (e.g., death, relapse, treatment failure). Such stratification, relevant across a variety of application domains, in this setting can support clinical decision-making, facilitate disease subtyping, and advance precision-medicine initiatives \citep{collins2015new}.
Nonetheless, patients are often intrinsically dependent, as they may share group membership -- such as being treated in the same hospital -- inducing correlations in outcomes and additional variability that must be accounted for in the analysis, as they can influence patients' risk profiles in complex ways.

This scenario is common in healthcare applications and is exemplified by the project motivating of our work, \textit{Enhance-Heart}, a study conducted in the Northern Italy regional district (Lombardy region), a region severely affected during the early stages of the COVID-19 pandemic due to an uncontrolled spread of SARS-CoV-2 infections \citep{alteri2021genomic}. 
This study focuses on patients with heart failure (HF) and aims, among other objectives, to investigate the complex relationship between patients' clinical profiles and COVID-19–related mortality, while also accounting for differences across hospitals.
Indeed, recent research has highlighted a significant interplay between heart failure and COVID-19 during the pandemic, with pre-existing heart failure complicating COVID-19 management and prognosis, and potentially leading to complications \citep{bader2021heart, adeghate2021mechanisms}.
Thus, identifying latent subgroups of HF patients hospitalized for COVID-19 while assessing how respiratory pathologies and overall healthcare condition influence mortality risk across patient clusters, represents a key step in characterizing outcomes in a time-to-event framework.
On the other hand, hospital-based resource availabilities, such as intensive care unit beds and general medicine/surgical beds, were found to be significantly associated with incidence rate of COVID-19 deaths \citep{janke2021analysis}.
Therefore, an accurate assessment of survival outcomes in this context requires modeling approaches capable of stratifying patient populations while accounting for differences across hospitals.

\medskip 

Traditional survival models often address heterogeneity arising from group-level dependence through a frailty term, that is a multiplicative factor in the hazard function, i.e. the risk of experiencing an event at a given time conditional on survival up to that point \citep{vaupel1979impact, clayton1978model}. This term can be specified in different ways, either parametrically -- most commonly using Gamma or Lognormal distributions -- or semiparametrically (see, e.g., \cite{abrahantes2007comparison, austin2017tutorial, balan2020tutorial}).
However, these models often assume homogeneity within groups and overlook latent similar patterns between patients induced by their risk of experiencing an event of interest at a given time. 
Conversely, unsupervised stratification methods can reveal hidden patient subgroups by leveraging similarities across individuals; however, their limited integration with time-to-event outcomes often leads to suboptimal predictive performance. For example, \cite{ahlqvist2018novel} performed a cluster analysis on six standardized clinical measures, identifying replicable diabetes subgroups with distinct prognoses and complication profiles.
Advancing toward survival-informed clustering, \cite{tosado2020clustering} proposed transforming the feature space according to each feature's association with survival (via martingale residuals) before applying clustering in this transformed space. Similarly, \cite{chen2016algorithm} introduced a recursive partitioning algorithm that derives prognostic cancer staging systems by maximizing survival separation between groups of patients. Shifting the focus towards prediction, \cite{chapfuwa2020survival} developed a Bayesian nonparametric framework that embeds subjects in a clustered latent space to enable accurate time-to-event predictions and subgroups with distinct risk profiles. Along related lines, \cite{manduchi2021deep} presented a variational-autoencoder-based survival clustering approach that jointly leverages patient features and time-to-event outcomes, while \cite{buginga2024clustering} proposed a nonparametric mixture-of-experts model in which each expert captures a distinct subgroup’s survival distribution, thereby uncovering heterogeneous risk patterns.
To the best of our knowledge, relatively few statistical models address the joint problem of clustering patients and modeling time-to-event outcomes while incorporating hierarchical structure. Among these, the cluster-weighted model of \cite{caldera2025cluster} represents a more recent contribution, in which a shared frailty model is fitted within each cluster and covariates are assumed to follow parametric distributions different in each cluster, with the overall joint density factorized via cluster-specific mixing weights. While this approach is promising, its reliance on strong distributional assumptions for the covariates may restrict its flexibility.

\medskip

To address this gap, we propose a novel model that extends parametric shared frailty models to account for group membership (e.g., patients treated in the same hospital) while simultaneously estimating both patient cluster membership -- defined by similar risk trajectories -- 
and hazard function parameters via penalized likelihood optimization, where the resulting solution is obtained through an iterative procedure.
Patient stratification is achieved through a modified version of spectral clustering approach following the strategies proposed in \cite{nie2014clustering, nie2016constrained, guo2015robust}, consisting in the adaptive estimation of a similarity matrix capturing patient-to-patient relationships representing the probability that patients are \textit{neighbors}, based on their estimated hazard risk, combining frailty and covariate effects.
At the end of the procedure, graph-theoretic principles ensure that the connected components of the resulting graph, represented by the learned similarity matrix, correspond exactly to the number of clusters.

Our model builds upon the framework proposed by \cite{liu2020supervised}, originally developed for cancer subtyping, which integrates supervised graph-based clustering into the modeling of multi-omic tumor data through penalization terms embedded in a \textit{partial} likelihood-based optimization scheme. Related extensions and applications of this approach have been explored in radiomics settings in \cite{cavinato2021recurrence, cavinato2023perspective}.

A key advantage of this framework is that patient clustering is embedded directly within the optimization procedure, so that cluster assignments are guided by underlying risk dynamics and reflect distinct event hazard profiles, making them interpretable in terms of event risk.
Building on this framework, our contribution is twofold:
\begin{enumerate}
    \item[(i)] we extend the model to account for group-level (hospital) dependence together with patient-level heterogeneity within a unified survival modeling framework;
    \item[(ii)] we propose a penalized \textit{full}-likelihood approach that embeds patient clustering directly within the hazard model, allowing similarity structures to be learned in a risk-driven manner and producing clusters that are interpretable in terms of event risk.
\end{enumerate}





\medskip

This paper is organized as follows.
Section \ref{sec:methodology} outlines the proposed methodology, detailing the notation and the estimation procedure.
In Section \ref{sec:simulation}, we present a simulation study designed to evaluate the performance of the proposed methodology.
In Section \ref{sec:casestudy}, we describe the \textit{Enhance-Heart} administrative database that motivated our work and report the main results obtained with our method.
Finally, Section \ref{sec:discussion} provides concluding remarks, discusses the strengths and limitations of the proposed approach, and outlines directions for future work.

The method, implemented through the software R (\cite{Rproject}), is openly accessible at \href{https://github.com/alessandragni/HazClust}{https://github.com/alessandragni/HazClust}.

\section{Methodology}
\label{sec:methodology}

In this section, we outline the methodology following three consecutive steps: a recap on notation for parametric frailty models (Subsection \ref{sec:notation}), the model description (Section \ref{sec:model}), the model optimization
(Section \ref{sec:optimization}) and some details related to the estimation procedure (Section \ref{sec:estimationdetails}).

\subsection{Notation}
\label{sec:notation}

Let the observed data be
$\mathcal{Z} = \{(y_{gi}, \delta_{gi}, \boldsymbol{x}_{gi}) \mid i = 1, \dots, n_g,\ g = 1, \dots, M\}$,
where units \( i = 1, \dots, n_g \) are nested within groups \( g = 1, \dots, M \), and the total number of units is \( N = \sum_{g=1}^M n_g \). For each unit, \( y_{gi} = \min(t_{gi}, C^*_{gi}) \) denotes the observed time to event or censoring, whichever comes first, and \( \delta_{gi} = \mathbb{I}(t_{gi} \leq C^*_{gi}) \) indicates whether the event was observed (\( \delta_{gi} = 1 \)) or censored (\( \delta_{gi} = 0 \)). The covariate vector \( \boldsymbol{x}_{gi} \in \mathbb{R}^{d \times 1} \) may include both group-level and unit-specific components. We assume independent and noninformative censoring \citep{munda2014adjusting}.
%
%
To model the hazard rate of subject $i$ in group $g$ at time $t$, accounting for the presence of heterogeneity across groups, among the possible approaches (e.g, fixed effects, stratified), we consider the shared frailty model \citep{duchateau2008frailty}
\begin{equation}
    h_{gi}(t \mid u_g) = h_0(t;\boldsymbol{\psi}) u_g \exp(\boldsymbol{x}_{gi}' \boldsymbol{\beta})
    \label{eq:frailtymodel}
\end{equation}
where $\boldsymbol{\beta} \in \mathbb{R}^{d\times1}$ is the vector of fixed-effect coefficients, the $u_g$'s, $g=1,\dots,M$, are the actual values of a random variable with probability density $f_U(\cdot \mid \theta)$, that is the frailty distribution, and $h_0(t; \boldsymbol{\psi})$ is the baseline hazard, assumed parametric, depending on the vector of parameters $\boldsymbol{\psi}$, and $'$ stands for the transpose.
The cumulative hazard then follows as
$H_{gi}(t \mid u_g) = u_g \exp(\boldsymbol{x}_{gi}' \boldsymbol{\beta}) H_0(t;\boldsymbol{\psi})$, with $H_0(t;\boldsymbol{\psi}) = \int_0^t h_0(s;\boldsymbol{\psi}) ds$.
Following computations in Appendix \ref{app:prooflikel}, the marginal loglikelihood can be expressed as
\begin{equation}
\label{eq:marginalloglik}
    \ell(\boldsymbol{\psi}, \boldsymbol{\beta}, \theta, \mathcal{Z}) = \sum_{g=1}^M \Bigg[ \bigg[\sum_{i=1}^{n_g} \delta_{gi} \left( \log(h_0(y_{gi};\boldsymbol{\psi})) + \boldsymbol{x}_{gi}' \boldsymbol{\beta} \right) \bigg] + \log \bigg[ (-1)^{d_g} \mathcal{L}^{(d_g)}\bigg( \sum_{i=1}^{n_g} H_0(y_{gi};\boldsymbol{\psi}) \exp(\boldsymbol{x}_{gi}' \boldsymbol{\beta}) \bigg)\bigg] \Bigg]
\end{equation}
where $d_g := \sum_{i=1}^{n_g} \delta_{gi}$ is the number of events in group $g$ and $(-1)^k\mathcal{L}^{(k)}(s) := (-1)^k \frac{d^k}{ds^k}\mathcal{L}(s) = \mathbb{E}[U^k\exp(-Us)]$, being $ \mathcal{L}(s)$ the Laplace transform, for $s\geq 0$ \citep{munda2014adjusting, klein1992semiparametric}.

Finally, it is known that the frailty term $u_g$ can be predicted by $\hat{u}_g = \mathbb{E}\big( U \mid \mathcal{Z}_g; \hat{\boldsymbol{\psi}}, \hat{\boldsymbol{\beta}}, \hat{\theta}\big)$,
being $\mathcal{Z}_g$ the observed data in group $g$ and
\begin{equation}
    \mathbb{E}\big( U \mid \mathcal{Z}_g; \boldsymbol{\beta}, \boldsymbol{\psi}, \theta\big) = - \frac{ \mathcal{L}^{(d_g+1)} \Big( \sum_{i=1}^{n_g} H_0(y_{gi};\boldsymbol{\psi}) \exp (\boldsymbol{x}_{gi}' \boldsymbol{\beta}) \Big)}{\mathcal{L}^{(d_g)} \Big( \sum_{i=1}^{n_g} H_0(y_{gi};\boldsymbol{\psi}) \exp (\boldsymbol{x}_{gi}' \boldsymbol{\beta}) \Big)} \, .
    \label{eq:conditionalexpectation}
\end{equation}

\subsection{Model}
\label{sec:model}

In a parametric setting, such as that addressed by \citep{munda2012parfm}, where both the frailty and the baseline hazard follow specified parametric distribution, the model parameters $\boldsymbol{\beta}, \boldsymbol{\psi}, \theta$ can be estimated by maximizing the loglikelihood in (\ref{eq:marginalloglik}), or minimizing the negative loglikelihood, as follows

\begin{equation}
\label{eq:problemsimple}
    \min_{\boldsymbol{\beta}, \boldsymbol{\psi}, \theta}  \quad -\ell(\boldsymbol{\psi}, \boldsymbol{\beta}, \theta, \mathcal{Z}) 
    \, .
\end{equation}

With the aim of performing clustering on the units while learning $\boldsymbol{\beta}, \boldsymbol{\psi}$ and $\theta$, let $\mathbf{S}\in \mathbb{R}^{N \times N}$ be the affinity matrix whose entries $\{s_{gi,hj}\}$ represent the similarity between the unit $i$ in group $g$ and the unit $j$ in group $h$. Specifically, we interpret each entry $\{s_{gi,hj}\}$ as the probability that unit $i$ in group $g$ has unit $j$ in group $h$ as a neighbor, imposing the constraint $\sum_{h=1}^M\sum_{j=1}^{n_h} s_{gi,hj} = 1$, $\forall g,i$ that ensures that the connection probabilities for each unit sum up to one.
By assuming that units have a higher probability of being neighbors if they are more similar, or equivalently have a lower distance $d_{gi,hj}(\boldsymbol{\psi}, \boldsymbol{\beta}, \theta, \mathcal{Z})$, we rewrite the optimization problem in Eq. (\ref{eq:problemsimple}) 
by further penalizing the loglikelihood for jointly estimating $\boldsymbol{\beta}, \boldsymbol{\psi}, \theta$ and $\mathbf{S}$:
\begin{align}
\label{eq:survivalfrailtyoptim_generald}
\min_{\boldsymbol{\beta}, \boldsymbol{\psi}, \theta, \mathbf{S}} \quad &   -\ell(\boldsymbol{\psi}, \boldsymbol{\beta}, \theta, \mathcal{Z}) 
+ \gamma \sum_{g,h=1}^{M} \sum_{i=1}^{n_g} \sum_{j=1}^{n_h}
\big( d_{gi,hj} (\boldsymbol{\beta}, \theta, \mathcal{Z}) \, s_{gi,hj} + \mu \; s_{gi,hj}^2  \big)\nonumber \\
\text{s.t.} \quad & \sum_{h=1}^{M} \sum_{j=1}^{n_h} s_{gi,hj} = 1, \quad \mathbf{s}_{gi} \geq 0, \quad \forall\, i = 1,\dots,n_g, \; g = 1,\dots,M
\end{align}
where $\mathbf{s}_{gi} \in \mathbb{R}^{1\times N}$ is the row of \textbf{S} corresponding to unit $i$ in group $g$, and the reported inequality is elementwise. 
The last two terms in Eq. (\ref{eq:survivalfrailtyoptim_generald}) follow the literature on clustering with adaptive neighbors, see for instance \cite{nie2014clustering}. These terms regulate the structure of neighborhood probabilities in $\mathbf{S}$: the first term encourages higher probabilities for nearby units, that, if alone leads to the limit case of 1 for the nearest neighbor and 0 elsewhere, while the quadratic term promotes smoother and more evenly distributed assignments, with the limit case assigning equal probability $1/N$ to all units \citep{nie2014clustering}. 
The hyperparameter $\gamma$ controls the strength of the penalization; when 
$\gamma = 0$, no penalization is applied, and consequently no clustering is induced, as the penalty and its associated constraints vanish.
In contrast, 
$\mu$ is a tunable parameter that modulates the influence of the similarity structure in the overall optimization. Its selection will be discussed in Sections~\ref{sec:optimization} and \ref{sec:estimationdetails}.

As we aim to learn $\mathbf{S}$ based on the 
on the similarity between the
risk of experiencing the event of interest for patient $i$ in group $g$, compared to patient $j$ in group $h$, that we quantify on the log-hazard scale,
as $\log h_{gi}(t \mid u_g) = \log h_0(t;\boldsymbol{\psi}) + \log u_g + \boldsymbol{x}_{gi}' \boldsymbol{\beta}$ 
and $\log h_{hj}(t \mid u_h)$, where the $\log h_0(t;\boldsymbol{\psi})$ shared across units cancels out in pairwise comparisons.
Thus, we define 
\begin{equation}
    d_{gi,hj} (\boldsymbol{\beta}, \theta, \mathcal{Z}) := 
\left\| (\boldsymbol{x}_{gi}' \boldsymbol{\beta} + \log u_g ) - (\boldsymbol{x}_{hj}' \boldsymbol{\beta} + \log u_h ) \right\|^2,
\label{eq:distance}
\end{equation}
where $\| \cdot \| $ as the Euclidean distance, and $u_g$ and $u_h$ are the frailty terms defined in (\ref{eq:conditionalexpectation}).
The resulting distance encourages the similarity graph to reflect 
frailty-adjusted risk alignment.

\medskip 

To perform clustering into $C$ clusters, we interpret $\mathbf{S}\in \mathbb{R}^{N \times N}$ as the adjacency matrix of a graph with the $N$ units as nodes. In this setting, we define the Laplacian matrix
$\mathbf{L}_\mathbf{S} := \mathbf{D}_\mathbf{S}-(\mathbf{S}'+\mathbf{S})/{2}$, where $\mathbf{D}_\mathbf{S}\in \mathbb{R}^{N \times N}$ is the diagonal degree matrix, where the $gi$-th entry is $\sum_{hj}(s_{gi,hj} + s_{hj,gi})/2$.
Following \cite{mohar1991laplacian}, the multiplicity $C$ of the zero eigenvalue of $\mathbf{L}_\mathbf{S}$ corresponds to the number of connected components in the graph with similarity matrix $\mathbf{S}$. Enforcing $rank(\mathbf{L}_\mathbf{S}) = N-C$ thus promotes a partition into $C$ clusters. Furthermore, as shown in \cite{nie2014clustering} and \cite{fan1949theorem}, this rank constraint can be handled equivalently and more easily by introducing a matrix $\mathbf{F} \in \mathbb{R}^{N \times C}$, given that 
\begin{equation}
\label{eq:traceequality}
    2 Tr(\mathbf{F}' \mathbf{L}_{\mathbf{S}} \mathbf{F}) = \sum_{g,h=1}^{M} \sum_{i=1}^{n_g} \sum_{j=1}^{n_h} \| \mathbf{f}_{gi} - \mathbf{f}_{hj} \|^2 s_{gi,hj},
\end{equation}
being $\mathbf{f}_{gi} \in \mathbb{R}^{C \times 1}$ the row of $\mathbf{F}$ corresponding to the belonging of unit $i$ in cluster $g$.

Combining (\ref{eq:marginalloglik}), (\ref{eq:survivalfrailtyoptim_generald}) and (\ref{eq:distance}), we get an optimization problem that reads
\begin{align}
\min_{\boldsymbol{\beta}, \boldsymbol{\psi}, \theta, \mathbf{S}, \mathbf{F}} \quad &   
- \Bigg\{ \sum_{g=1}^M \Bigg[ \bigg[\sum_{i=1}^{n_g} \delta_{gi} \left( \log(h_0(y_{gi};\boldsymbol{\psi})) + \boldsymbol{x}_{gi}' \boldsymbol{\beta} \right) \bigg] + \log \bigg[ (-1)^{d_g} \mathcal{L}^{(d_g)}\bigg( \sum_{i=1}^{n_g} H_0(y_{gi};\boldsymbol{\psi}) \exp(\boldsymbol{x}_{gi}' \boldsymbol{\beta}) \bigg)\bigg] \Bigg] \Bigg\}  \nonumber\\
 & 
+ \gamma \Bigg\{ \sum_{g,h=1}^{M} \sum_{i=1}^{n_g} \sum_{j=1}^{n_h}
\big[ 
\left\| (\boldsymbol{x}_{gi}' \boldsymbol{\beta} + \log u_g ) - (\boldsymbol{x}_{hj}' \boldsymbol{\beta} + \log u_h ) \right\|^2 
\, s_{gi,hj} + \mu \; s_{gi,hj}^2 \big] \nonumber  \\
& 
 + 2\lambda Tr(\mathbf{F}' \mathbf{L}_{\mathbf{S}} \mathbf{F}) \Bigg\} \nonumber \\
\text{s.t.} \quad & \sum_{h=1}^{M} \sum_{j=1}^{n_h} s_{gi,hj} = 1, \quad \mathbf{s}_{gi} \geq 0, \quad \forall\, i = 1,\dots,n_g, \; g = 1,\dots,M, \quad \mathbf{F}' \mathbf{F} = \mathbf{I}, \, \mathbf{F} \in \mathbb{R}^{N \times C} 
\label{eq:problemcomplete}
\end{align}
where 
$$ \log u_g = \log \Big[ (-1)^{d_g+1}\mathcal{L}^{(d_g+1)} \Big( \sum_{i=1}^{n_g} H_0(y_{gi};\boldsymbol{\psi}) \exp (\boldsymbol{x}_{gi}' \boldsymbol{\beta}) \Big) \Big] - \log \Big[{(-1)^{d_g} \mathcal{L}^{(d_g)} \Big( \sum_{i=1}^{n_g} H_0(y_{gi};\boldsymbol{\psi}) \exp (\boldsymbol{x}_{gi}' \boldsymbol{\beta}) \Big)} \Big] \, ,$$ 
which follows from (\ref{eq:conditionalexpectation}) by noting that $-1 = {(-1)^{d_g+1}}/{(-1)^{d_g}}
$ and applying a logarithmic property.

\subsection{Optimization procedure}
\label{sec:optimization}

To solve the optimization problem in Eq.~(\ref{eq:problemcomplete}), we adopt a block coordinate descent approach: at each iteration, we update one block of variables while keeping the others fixed, until a convergence criterion is met (see Subsection~\ref{sec:estimationdetails} for details).
The procedure alternates among the following three steps:

\begin{enumerate}
    \item \textbf{Fix $\mathbf{S}$, $\boldsymbol{\beta}, \boldsymbol{\psi}, \theta$ and update $\mathbf{F}$}

With $\mathbf{S}$, $\boldsymbol{\beta}$, $\boldsymbol{\psi}$, and $\theta$ fixed, the update of $\mathbf{F}$ reduces to the following spectral optimization problem \citep{nie2014clustering}:
\begin{align*}
    \min_{\mathbf{F}} \quad 
& 2\lambda Tr(\mathbf{F}^T \mathbf{L}_\mathbf{S} \mathbf{F}) \\
\text{s.t.} \quad & \mathbf{F}^T \mathbf{F} = \mathbf{I}, \mathbf{F} \in \mathbb{R}^{N \times C}
\end{align*}

The factor $2\lambda$ does not affect the solution and can be omitted.  
The optimal $\mathbf{F}$ is obtained by taking the $C$ eigenvectors of the Laplacian $\mathbf{L}_\mathbf{S}$ associated with its $C$ smallest eigenvalues.  
This step requires the number of clusters $C$ and the current estimate of $\mathbf{S}$.


    \item \textbf{Fix $\mathbf{F}, \mathbf{S}$ and update $\boldsymbol{\beta}, \boldsymbol{\psi}, \theta$} 

    With $\mathbf{F}$ and $\mathbf{S}$ fixed, the optimization over $(\boldsymbol{\beta}, \boldsymbol{\psi}, \theta)$ reduces to a survival modeling problem with an additional similarity-based regularization term.  
    The optimization problem can be expressed as:
    \begin{align*}
    \min_{\boldsymbol{\beta}, \boldsymbol{\psi}, \theta } \quad &   -\ell(\boldsymbol{\beta}, \boldsymbol{w}, \theta, \mathcal{Z})  
    + \gamma \sum_{g,h=1}^{M} \sum_{i=1}^{n_g} \sum_{j=1}^{n_h}
\left(
\left\| \boldsymbol{x}_{gi}' \boldsymbol{\beta} + \log u_g - \boldsymbol{x}_{hj}' \boldsymbol{\beta} - \log u_h \right\|^2
\right)
s_{gi,hj} 
\end{align*}

We estimate the survival model parameters relying on maximum likelihood optimization with parametric frailty, following \cite{munda2012parfm}, and extend this framework with the similarity-based penalty.  
This step requires $\gamma$, the current similarity matrix $\mathbf{S}$, the dataset and model specification.

\item \textbf{Fix $\mathbf{F}, \boldsymbol{\beta}, \boldsymbol{\psi}, \theta$ and update $\mathbf{S}$} 

With $\mathbf{F}$, $\boldsymbol{\beta}$, $\boldsymbol{\psi}$, and $\theta$ fixed, the update of $\mathbf{S}$ is obtained by solving
\begin{align*}
    \min_{\mathbf{S}} \quad 
& \gamma \Bigg\{ \sum_{g,h=1}^{M} \sum_{i=1}^{n_g} \sum_{j=1}^{n_h}
\left[ 
\left\| \boldsymbol{x}_{gi}' \boldsymbol{\beta} + \log u_g - \boldsymbol{x}_{hj}' \boldsymbol{\beta} - \log u_h \right\|^2
s_{gi,hj}
+ \mu s_{gi,hj}^2 \right] \\
& + 2\lambda Tr(\mathbf{F}' \mathbf{L}_{\mathbf{S}} \mathbf{F}) \Bigg\} \\
\text{s.t.} \quad & \sum_{h=1}^{M} \sum_{j=1}^{n_h} s_{gi,hj} = 1, \quad 0 \leq \mathbf{s}_{gi} \leq 1, \quad \forall\, i = 1,\dots,n_g, \; g = 1,\dots,M
\end{align*}
Using relation~(\ref{eq:traceequality}) and the separability of the problem across units, this reduces to an independent optimization for each $\mathbf{s}_{gi}$:
\begin{align}
    \min_{\mathbf{s}_{gi}} \quad 
& \sum_{h=1}^{M} \sum_{j=1}^{n_h}
\left(
\left\| \boldsymbol{x}_{gi}' \boldsymbol{\beta} + \log u_g - \boldsymbol{x}_{hj}' \boldsymbol{\beta} - \log u_h \right\|^2
+ \mu s_{gi,hj} + \lambda \| \mathbf{f}_{gi} - \mathbf{f}_{hj} \|^2
\right ) s_{gi,hj} \nonumber \\
\text{s.t.} \quad & \sum_{h=1}^{M} \sum_{j=1}^{n_h} s_{gi,hj} = 1, \quad \mathbf{s}_{gi} \geq 0 
\label{eq:step3longform}
\end{align}

As shown in Appendix~\ref{app:proofKKT}, this problem admits a closed-form solution for each row $\mathbf{s}_{gi}$:
    $$s_{gi,hj} = \max\big\{0, \alpha_{gi} - w_{gi,hj}/(2\mu_{gi}) \big\}$$
subject to $\mathbf{s}_{gi}' \mathbf{1} = 1$, where $\alpha_{gi} \geq 0$ and and $\mu_{gi}$ may differ across units. Here, $\boldsymbol{w}_{gi} \in \mathbb{R}^{N \times1}$ and
$$w_{gi, hj} := 
\left\| \boldsymbol{x}_{gi}' \boldsymbol{\beta} + \log u_g - \boldsymbol{x}_{hj}' \boldsymbol{\beta} - \log u_h \right\|^2
+ \lambda \| \mathbf{f}_{gi} - \mathbf{f}_{hj} \|^2.$$ 

By enforcing sparsity on $\mathbf{s}_{gi}$, i.e., allowing only the $k$ nearest neighbors to connect unit $i$ in group $g$, 
following \cite{nie2014clustering, guo2015robust} we impose additional constraints that make the problem well-posed while also reducing computational cost. 
Let $\hat{\boldsymbol{w}}_{gi} = (\hat{w}_{gi,1}, \ldots, \hat{w}_{gi,N-1})$ denote the sorted distances (excluding the self-distance $w_{gi,gi}$). 
Then the optimal ordered similarity vector $\hat{\mathbf{s}}_{gi}$ is constrained to have exactly $k$ nonzero entries:
\begin{enumerate}[(i)]
    \item $\hat{s}_{gi,p} > 0$ for $p = 1, \dots, k$, corresponding to the $k$ nearest neighbors;
     \item $\hat{s}_{gi,p} = 0$ for $p = k+1, \dots, N$, including self-similarity. 
\end{enumerate}
Under these conditions, and the constraint $\mathbf{s}_{gi}' \mathbf{1} = 1$, as shown in Appendix \ref{app:getalphamu}, we obtain
\begin{equation}
    \alpha_{gi} = \frac{1}{k} + \frac{1}{2 k \mu_{gi}} \sum_{p=1}^{k} \hat{w}_{gi,p} \quad \text{and} \quad \mu_{gi} = \frac{k}{2} \hat{w}_{gi,(k+1)} - \frac{1}{2} \sum_{p=1}^{k} \hat{w}_{gi,p}.
    \label{eq:alpha_mu}
\end{equation}

Finally, once all rows are updated, we set $\mu = \sum_{g,i} \mu_{gi} / N$.

This step requires $\lambda$, $k$, the current estimate of $\mathbf{F}$, the dataset and the model specification.

\end{enumerate}

\subsection{Estimation details}
\label{sec:estimationdetails}

This section provides practical details and technicalities on the estimation and implementation of the proposed algorithm.

\medskip
\textbf{Initialization.} If no prior information is provided, the similarity matrix $\mathbf{S}$ is initialized with zeros on the diagonal and 
$1/(N-1)$ for all off-diagonal entries, so each row sums to 1.
An initial similarity matrix can alternatively be supplied via \texttt{S0}.
The penalty parameter $\lambda$ is set to $\lambda_0 = 1000$ by default, as it is advised in the literature to set it large enough \citep{nie2014clustering}, but may be specified by the user through the parameter \texttt{lambda0}. After the first update of $\mathbf{S}$, $\lambda$ is reset to the value of $\mu$ obtained from the closed-form solution, and then adaptively adjusted: it is increased (multiplied by 1.2) if the number of connected components in $\mathbf{S}$ is smaller than $C$, or decreased (divided by 1.2) otherwise \citep{nie2017multi}.
    
The regression and frailty parameters are initialized either based on user input via the \texttt{inip} and \texttt{iniFpar} options, or using default values as done in the \texttt{parfm} R package \citep{munda2012parfm}.
As in this package, our software supports a range of parametric baseline hazard distributions
(Weibull, exponential, Gompertz, log-normal, \dots), as well as several frailty
distributions (gamma, log-normal, positive stable, among others).  

\medskip
\textbf{Convergence.} 
Convergence is declared when the algorithm stabilizes both the similarity matrix and the model fit. Specifically, the procedure stops when all of the following are satisfied: (i) the mean absolute change in \textbf{S} is below \texttt{tolS}, indicating that the graph structure has stabilized; (ii) the improvement in log-likelihood is below \texttt{tolll}, showing that the statistical fit has plateaued; and (iii) the number of connected components in \textbf{S} equals the target $C$. These criteria are evaluated jointly, and the algorithm terminates when all are met or when a maximum of \texttt{maxit} iterations is reached. Additionally, the inner optimization in step 2, which riles on \texttt{parfm}, can be limited to \texttt{maxitparfm} iterations.
Defaults are set to \texttt{tolS} = $10^{-4}$, \texttt{tolll} = $10^{-3}$, \texttt{maxit} = 500, \texttt{maxitparfm} = 500.

\medskip
\textbf{Choice of $k$ and its relation to $C$.}
The number of nearest neighbors $k$ controls the sparsity of the initial similarity matrix $\mathbf{S}$ and thereby the density of the corresponding graph. Indeed, each row contains exactly $k$ nonzero entries, enforcing a fixed local neighborhood structure and excluding self-similarity (the diagonal of $\mathbf{S}$ is set to zero).

The choice of $k$ strongly influences how readily the constructed graph can achieve the target of $C$ connected compontents. If $k$ is too small, the graph may fragment into more than $C$ components, whereas a large $k$ yields a dense graph that tends to collapse structure into fewer than $C$ components. Our algorithm compensates for this interaction by adaptively tuning the penalty parameter $\lambda$ to enforce convergence to the prescribed number~$C$. However, in practice, $k$ should be selected to guarantee basic connectivity without sacrificing sparsity, and needs to be carefully tuned depending on the application. A common heuristic is to let $k$ grow moderately with the sample size.

\medskip
\textbf{Model selection and choice of $C$.} 
To assess the quality of the clustering solution, we employ the silhouette coefficient \citep{kaufman2009finding}.
Given a unit $i$, let $a(i)$ denote its average distance to all other units in the same
cluster, and let $b(i)$ be the minimum average distance from $i$ to any other cluster.
We recall that the silhouette value is defined as
\[
s(i) = \frac{b(i) - a(i)}{\max\{a(i),\, b(i)\}},
\quad s(i)\in [-1,1], \]
with higher values indicating better separation.
In our setting, distances for unit $i$ with respect to units $j$ are computed as $\sqrt{d_{gi,hj} (\boldsymbol{\beta}, \theta, \mathcal{Z})}$ in Eq. (\ref{eq:distance}), as it incorporates the covariate effects and the estimated frailty contributions. 
Group indices (such as $g$ for $i$ and $h$ for $j$) are omitted here since they play no role in the silhouette calculation.
We then assess the overall clustering quality using the mean silhouette, 
and select the optimal clustering configuration as the one with the maximum mean silhouette, thereby achieving the best separation with respect to the fitted survival model.

\medskip
\textbf{Software.} The algorithm is implemented in \textsf{R} and is available in at the following repository
\href{https://github.com/alessandragni/HazClust}{https://github.com/alessandragni/HazClust}.
The core survival and frailty estimation routines
rely on the \texttt{parfm} package \citep{munda2012parfm}, which provides
maximum likelihood estimation for parametric frailty models. Our contribution
extends \texttt{parfm} by embedding it within an iterative graph-based clustering framework.

\section{Simulation Study}
\label{sec:simulation}

In this section, we present a simulation study to assess the performance of the proposed methodology. 
%
%
Our main focus is on evaluating how well the method recovers the underlying cluster structure across different levels of penalization. We also investigate the behavior of the parameter estimates under varying penalization strengths. Additionally, when the latent structure contains $C$ clusters, we assess how accurately the method identifies the true clustering compared to alternative choices of $C$, with the mean silhouette providing a useful overall indicator of clustering quality.

\subsection{Data generating process}
\label{sec:DGP}


We generate data by inducing the presence of $M = 10$ groups, each containing $n_g = 50$ ($g=1,\dots,M$), yielding a total sample size of $N = 500$. 
To induce dependence among units within the same group, we introduce a group-level frailty $u_g$, drawn independently for each group from a Gamma distribution $u_g \sim \text{Gamma}(1/\theta, 1/\theta)$, with $\theta=0.5$ so that $\mathbb{E}[u_g]=1$ and $\text{Var}(u_g) = \theta$.
Keeping the frailty variance small ensures that cluster-specific effects remain the dominant source of heterogeneity.
Further elaboration of this aspect will be provided in the final Discussion.

Within each group $g$, every unit $i$ is randomly assigned to one of the $C=3$ latent clusters, denoted $z_{gi} \in \{1,2,3\}$.
Each unit is associated with a covariate vector $\boldsymbol{x}_{gi}$ and a linear predictor
$\eta_{gi} = \boldsymbol{x}_{gi}' \boldsymbol{\beta} + \log u_g $ 
where $\boldsymbol{\beta}$ is a vector of regression coefficients,
so that, conditional on the cluster assignment $z_{gi}$ and frailty $u_g$, the linear predictor is modeled as 
$$ \eta_{gi} \mid z_{gi} = c, u_g \sim \mathcal{N}(\mu_{c}, \sigma^2), \quad c=1,2,3$$
where $[\mu_1, \mu_2, \mu_3] =  [-8,0,8]$ and $\sigma=1$.
This specification generates clearly separated distributions of $\eta_{gi}$ across clusters,
maximizing the clustering structure as measured in Eq. (\ref{eq:distance}).
Finally, to preserve the intended simulation structure across clusters, we construct a single observed covariate $x_{gi}$ as $x_{gi} = \big( \eta_{gi} - \log u_g \big) / {\beta}$, setting $\beta = \log(2)$.

Survival times are generated under a Weibull baseline hazard, parameterized 
by $\boldsymbol{\psi} = [\rho, \psi]$ through $H_0(t) = \xi^\rho t^\rho$ ($\rho, \xi >0$), with $\rho = 2.5$ and $\xi = 0.01$.
The corresponding inverse cumulative hazard is $H_0^{-1}(s) = s^{1/\rho}/\xi$.
Event times are generated using inverse transform sampling: we draw $V_{gi} \sim \mathrm{Uniform}(0,1)$ and set $t_{gi} = H_0^{-1}\big(-\log(V_{gi})/\exp(\eta_{gi})\big)$.

Censoring is imposed using one of two possible mechanisms:
(i) \textit{administrative} censoring at a fixed time of 100, and (ii) \textit{Gaussian (normal)} censoring with mean 130 and standard deviation 15,
analogous to the default option of the 
\texttt{genfrail} function of the \texttt{frailtySurv} \textsf{R} package \citep{monaco2018general}.

Under this data generation mechanims, the Kaplan–Meier curves of each of the generated data highlight distinct patterns across the three clusters: one maintains high survival, another declines gradually, and a third drops sharply early on.

\medskip

Following the data generating process just described, 
we set $$\gamma \in \{0, 10^{-6}, 10^{-4}, 10^{-3}, 10^{-2}, 0.1, 0.2, 0.4\}.$$
The case $\gamma = 0$ corresponds to absence of penalization, so no clustering is performed, and just a parametric frailty model as in \texttt{parfm} is fitted.
When $\gamma > 0$, we set $C \in \{2,3,4,5\}$ and $k = \{20, 50\}$. 
For each setting, we replicate the simulation $B = 100$ times.
In each replication, the same simulated dataset is used, while the input parameters of the model are varied to assess its behavior under different configurations.
Furthermore, we set \texttt{tolS} = $10^{-2}$, \texttt{tolll} = $1$ and \texttt{maxit} = $200$.

\medskip

\subsection{Results}

We first focus on the simulations with $C=3$ and $\gamma > 0$.
For both $k = 20$ and $k = 50$, the method consistently identifies three clusters, with only minor differences in the $\gamma$ thresholds at which performance begins to decline. Under \textit{administrative} censoring, the method identifies the three clusters with 100\% frequency for $\gamma \le 0.1$. For larger $\gamma$, the identification rate decreases slightly, reaching a minimum of 95\% at $\gamma = 0.4$. 
Overall, convergence was achieved after an average of 17.49 iterations.
Under \textit{normal} censoring, perfect identification is maintained for $\gamma \le 10^{-3}$, after which the rate gradually slightly declines, reaching a minimum of 98\% at $\gamma = 0.4$.
In this second case, the algorithm required on average 24.49 iterations to converge.

Considering now only the cases in which three clusters were correctly identified, clustering performance is evaluated using accuracy and the adjusted Rand index (ARI), with the true cluster memberships of each unit $i$ in group $g$ known from the simulation design.
Accuracy measures the proportion of observations whose predicted cluster labels match the true labels after optimal label alignment.
ARI quantifies pairwise agreement corrected for chance, ranging in $[-1,1]$, with 1 indicating perfect recovery, 0 random clustering, and negative values worse-than-chance agreement.
Results are reported in Table~\ref{tab:AccuracyandARI} across different model configurations, that is, varying censoring mechanism, $k$, and $\gamma$.

\begin{table}[!htbp] 
\centering
\footnotesize
\begin{tabular}{lll ccc ccc}
\toprule
\textbf{Censoring} & $k$ & $\gamma$
    & \multicolumn{3}{c}{\textbf{Accuracy}} 
    & \multicolumn{3}{c}{\textbf{ARI}} \\
\cmidrule(lr){4-6} \cmidrule(lr){7-9}
 & & & Mean & Median & SD & Mean & Median & SD \\
\toprule

\multirow{14}{*}{(i) Administrative} 
  & \multirow{7}{*}{20}
    & $10^{-6}$ & 0.971 & 1.000  & 0.089 & 0.942 & 1.000 & 0.161\\
  &  & $10^{-4}$ & 0.971 & 1.000  & 0.089 & 0.942 & 1.000 & 0.161 \\
  &  & $10^{-3}$ & \textbf{0.975} & \textbf{1.000} & \textbf{0.081} & \textbf{0.950} & \textbf{1.000} & \textbf{0.147} \\
  &  & $10^{-2}$ & 0.957 & 1.000 & 0.082 & 0.907 &  1.000 & 0.172 \\
  &  & $0.1$ & 0.950 & 1.000  & 0.083 & 0.888 & 1.000 & 0.179 \\
  &  & $0.2$ & 0.930 & 1.000  & 0.115 & 0.860 & 1.000 & 0.197 \\
  &  & $0.4$ & 0.892 & 0.928  & 0.141 & 0.788 & 0.819   & 0.242 \\

\cmidrule(lr){2-9}

  &  \multirow{7}{*}{50}
   & $10^{-6}$ & \textbf{1.000} & \textbf{1.000} & \textbf{0.001} & \textbf{0.999} & \textbf{1.000} & \textbf{0.003}\\
  &  & $10^{-4}$ & \textbf{1.000} & \textbf{1.000} & \textbf{0.001} & \textbf{0.999} & \textbf{1.000} & \textbf{0.003}\\
  &  & $10^{-3}$ & \textbf{1.000} & \textbf{1.000} & \textbf{0.001} & \textbf{0.999} & \textbf{1.000} & \textbf{0.003}\\
  &  & $10^{-2}$ & 1.000 & 1.000 & 0.001 & 0.999 & 1.000 & 0.004\\
  &  & $0.1$ & 0.996 & 1.000 & 0.023 & 0.980 & 1.000 & 0.054\\
  &  & $0.2$ & 0.945 & 0.998 & 0.120 & 0.897 & 0.994 & 0.199\\
  &  & $0.4$ & 0.420 & 0.390 & 0.109 & 0.060 & 0.019  & 0.159\\

\midrule
\multirow{14}{*}{(ii) Normal} 
  & \multirow{7}{*}{20}
    & $10^{-6}$ & 0.902 & 0.998 & 0.142 & 0.810 & 0.994   & 0.246\\
  &  & $10^{-4}$ & 0.904 & 0.998 & 0.138 & 0.810 & 0.994  & 0.251\\
  &  & $10^{-3}$ & 0.891 & 0.970 & 0.141 & 0.788 & 0.918 &  0.243\\
  &  & $10^{-2}$ & 0.896 & 0.998 & 0.149 & 0.807 & 0.994 & 0.239\\
  &  & $0.1$ & 0.913 & 1.000 & 0.117 & 0.820 & 1.000 &  0.224\\
  &  & $0.2$ & \textbf{0.922} & 0.998 & \textbf{0.097} & \textbf{0.831} & 0.994 &  \textbf{0.195} \\
  &  & $0.4$ & 0.870 & 0.894 & 0.145 & 0.752 & 0.732 &   0.237\\

\cmidrule(lr){2-9}

  & \multirow{7}{*}{50}
    & $10^{-6}$ & 0.990 & 1.000 & 0.040 & 0.976 & 1.000   & 0.094\\
  &  & $10^{-4}$ & 0.990 & 1.000 & 0.040 & 0.977 & 1.000  & 0.094\\
  &  & $10^{-3}$ & 0.990 & 1.000 & 0.041 & 0.976 & 1.000 &  0.096\\
  &  & $10^{-2}$ & \textbf{0.998} & \textbf{1.000} & \textbf{0.016} & \textbf{0.994} & \textbf{1.000}  & \textbf{0.034}\\
  &  & $0.1$ & 0.994 & 1.000 & 0.027 & 0.986 & 1.000   & 0.060 \\
  &  & $0.2$ & 0.964 & 0.998 & 0.089 & 0.922 & 0.994 &   0.183\\
  &  & $0.4$ & 0.417 & 0.386 & 0.102 & 0.056 & 0.015   & 0.148 \\
\bottomrule
\end{tabular}
\caption{\small Empirical means, medians, and standard deviations of \textit{accuracy} and \textit{ARI} over $B=100$ replications, evaluated under two censoring mechanisms and across varying values of $k$ and $\gamma$, setting $C=3$. The best estimates in each setting are highlighted in bold.}
\label{tab:AccuracyandARI}
\end{table}

Clustering performance is generally very high, especially for larger $k$, with accuracy and ARI frequently approaching 1.
Specifically, the median remains close to 1 across most settings, reflecting the skewed distribution of performance measures near the upper bound; the mean, in contrast, shows the gradual rise and the sharp drop more clearly: performance is slightly lower for very small $\gamma$, reaches its maximum at moderate $\gamma$ of the chosen set, and then declines sharply once $\gamma$ reaches 0.4, as penalization becomes much stronger.
When accuracy and ARI are highest, their standard deviations are extremely small, indicating highly stable performance across replications.

\medskip

\begin{figure}[ht]
\centering
\includegraphics[width=\linewidth]{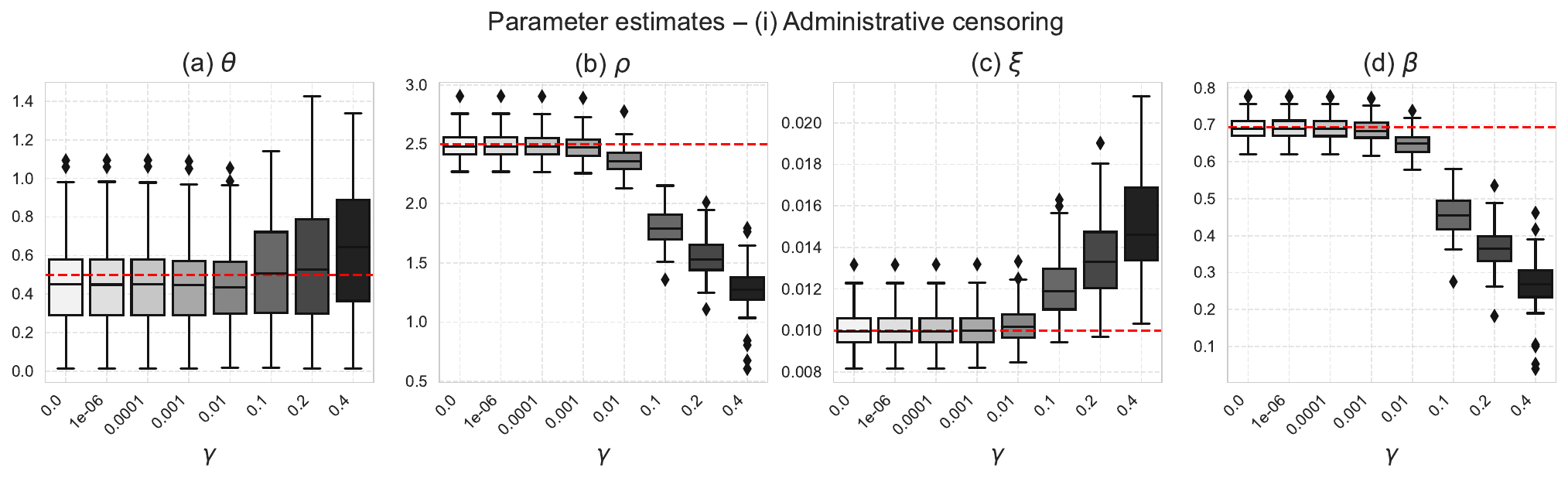}
\caption{Boxplots of parameter estimates ($\theta, \rho, \xi, \beta$) across penalization levels $\gamma$ (x-axis) under \textit{administrative} censoring for $k=20$ and $C=3$. The true parameter values are shown as a piecewise-constant horizontal line.}
\label{fig:all_estimates_admin}
\end{figure}

\begin{figure}[ht]
\centering
\includegraphics[width=\linewidth]{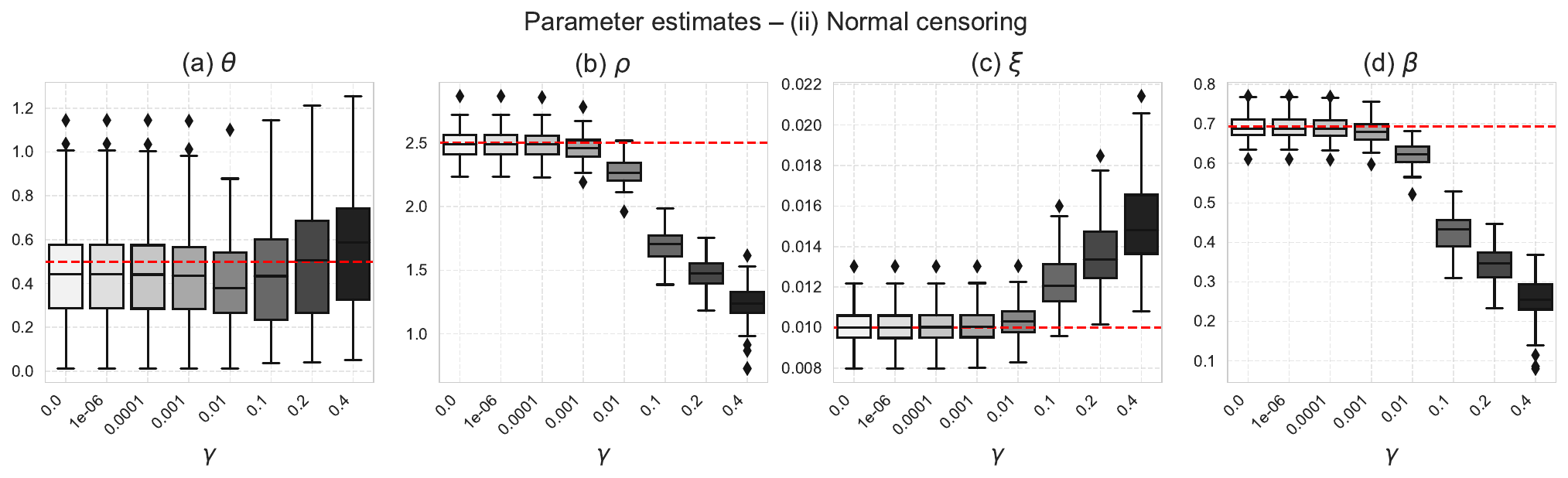}
\caption{Boxplots of parameter estimates ($\theta, \rho, \xi, \beta$) across penalization levels $\gamma$ (x-axis) under \textit{normal} censoring for $k=20$ and $C=3$. The true parameter values are shown as a piecewise-constant horizontal line.}
\label{fig:all_estimates_norm}
\end{figure}

Focusing still on the case $C=3$, we now restrict attention to $k=20$ to ease visualisation and reduce the number of scenarios presented.
In this setting, we examine the estimates of the unknown parameters in the shared frailty model, $[\theta, \rho, \xi, \beta]$ (we recall that in Section \ref{sec:methodology}, we named $\boldsymbol{\psi} = [\rho, \xi]$ as the Weibull distribution is parameterized by two parameters). For completeness, we also include $\gamma = 0$, which allows us to visualise the parameter estimates obtained in the absence of penalisation.
Figures \ref{fig:all_estimates_admin} and \ref{fig:all_estimates_norm} display boxplots of the estimated parameters across different values of $\gamma$, 
for administrative and normal censoring, respectively. 
In these boxplots, the true parameter values are indicated by a dashed horizontal line, allowing a visual assessment of bias.
As expected, the bias in the estimates generally increases with $\gamma$, represented on the x-axis. 
In particular, the parameter 
$\theta$ in plot (a), representing the variance of the frailty, shows a slight underestimation even in the unpenalized case ($\gamma = 0$, first boxplot), a pattern that is consistent with previous studies 
relying on the implementation of \texttt{parfm} package \citep{munda2012parfm}. 
Moreover, most parameters show relatively stable behaviour in terms of variability across different $\gamma$ values, indicating that penalization has a limited effect on their spread. 
The only exception is $\xi$, the second parameter of the Weibull, which appears to be the most affected, with its variability increasing noticeably as $\gamma$ increases.

To evaluate how penalization affects the quality of the parameter estimates, Table~\ref{tab:MSE_Variance} reports their variance (Var) and the mean squared error normalized by variance (MSE/Var) across all values of $\gamma$ and both censoring schemes. Appendix~\ref{app:msevariance} provides a detailed interpretation of these results.

\medskip

We now turn to analysing the results obtained varying $C$.
In Table \ref{tab:Silhouette}, we analyse the  empirical distribution of the mean silhouette values for varying values of $C$ under the two censoring mechanisms, and a specific choice of $k = 20$ and $\gamma = 10^{-4}$. The results indicate that 
$C=3$ yields the highest average and median mean silhouette values in both scenarios, suggesting that this choice provides the best overall cluster separation.
Although the silhouette values for $C=3$ exhibit slightly greater variability, the overall separation remains substantially stronger.
Moreover, convergence to the selected number of clusters within the prescribed number of iterations is always reached at $C=3$ across all replications, while it is not the case for all the other choices.

\begin{table}[!htbp] 
\centering
\footnotesize
\begin{tabular}{llccccc}
\toprule
\textbf{Censoring} & $C$ & \multicolumn{3}{c}{\textbf{Mean silhouette}} & \textbf{Convergence} \\
\cmidrule(lr){3-5}
& & Mean & Median & SD & \%\\
\toprule
\multirow{4}{*}{(i) Administrative} 
  & 2 &  0.660  &  0.670 & 0.046 & 67\\
  & \textbf{3} &  \textbf{0.792}  &  \textbf{0.832} & \textbf{0.116} & \textbf{100}\\
  & 4 &  0.661 & 0.718 &  0.121 & 100\\
  & 5 &  0.566 & 0.612 &  0.113 & 100\\

\midrule
\multirow{4}{*}{(ii) Normal} 
  & 2 &  0.618 & 0.649 & 0.082  & 93\\ 
  & \textbf{3} & \textbf{0.670}  &  \textbf{0.789} &  \textbf{0.182} &  \textbf{100}\\ 
  & 4 & 0.541  &  0.577 &  0.173 & 99\\ 
  & 5 & 0.466  &  0.484 &  0.145 & 100\\ 

\bottomrule
\end{tabular}
\caption{\small Summary of the empirical distribution of the mean silhouette values over $B=100$ 
replications under two censoring mechanisms, for $k=20$ and $\gamma = 10^{-4}$. For each value of $C$, the table reports 
the mean, median, standard deviation, and the proportion of runs that converged to the selected $C$ within \texttt{maxit} iterations. The estimates related to $C=3$ are highlighted in bold.}
\label{tab:Silhouette}
\end{table}

\section{Case Study}
\label{sec:casestudy}

This section applies the proposed methodology to the \textit{Enhance-Heart} dataset, a large administrative dataset linked with the 
Covid-19 registry\footnote{\href{https://www.dati.lombardia.it/}{https://www.dati.lombardia.it/}} of Lombardy Region (RL in the following).

\subsection{Cohort selection}

Sourcing from the administrative dataset related to the \textit{Enhance-Heart} project, we select those patients that between \textit{January 1, 2018}, and \textit{December 31, 2020} were diagnosed with HF, that is, hospitalized or visited ER under DRG code 127 (\lq Heart Failure and Shock') according to the RL MS-DRG v40 system, thus including both primary/secondary diagnoses of HF (ICD-9-CM: 428.xx) and related conditions (e.g. ICD-9-CM: 402.01, 402.11, 402.91).
Among these, we retain those hospitalized for COVID-19 according to a binary flag provided in the database, between \textit{31 January 2020} (start of the national COVID-19 state of emergency in Italy) and \textit{18 June 2021} (implementation of the Green Pass regulations governing reopening).
For those patients, outcomes are assessed over a 90-day follow-up period; death within this window is considered the event, while patients surviving beyond or dying later are regarded as censored (55.25\% of the sample).

The analysis is restricted to hospitals with a minimum of 50 patients hospitalized, 
as smaller hospitals may exhibit different organizational conditions and characteristics, and to ensure more robust results for the frailty-related estimates. 
We further exclude patients who required hospital transfers for medical purposes or who had multiple episodes of infection. 
The final cohort includes $N = 3086$ HF patients hospitalized for COVID-19 in $J = 32$ different hospitals in the RL.

The time-to-event outcome is defined by the pair (\texttt{Time}, \texttt{Status}), where \texttt{Time} denotes days until death within the 90-day observation window or censoring, and \texttt{Status} indicates event occurrence through a binary variable.
Patient's covariates include \(\texttt{Age}\) at admission, \(\texttt{Gender}\), and
Modified Multisource Comorbidity Score (\texttt{ModMCS}), a measure of overall comorbidity burden unrelated to the primary diagnosis and a proxy for general health status, 
with higher values reflecting greater comorbidity. It was introduced by \cite{corrao2017developing} and adapted by \cite{caldera2025uncovering} to exclude respiratory diseases relevant to COVID-19.
Indeed, we have considered separately the variable \(\texttt{Resp}\) defining it as a binary indicator that equals 1 when more than one respiratory condition - among chronic obstructive pulmonary disease, bronchitis, pneumonia, or respiratory failure - is present.
A detailed description and summary statistics of the study variables are reported in Table~\ref{Descriptive_variables_initial}.

 \begin{table}[!ht]
\caption{\small List, description and summary statistics of selected variables in the RL database.}
\label{Descriptive_variables_initial}
\renewcommand{\arraystretch}{1.5} 
\footnotesize
\begin{tabularx}{\textwidth}{@{}lXlX@{}}
\toprule
Variable 
& Description
& Type 
& Summary statistics\\
\midrule
\texttt{Status}& Binary indicator for death, equal to 1 if the event occurs within the observation period, 0 otherwise. &Categorical & 44.75\% for 1 (death) and 55.25\% for 0 (alive). \\
\texttt{Time}  & Either the survival time (in days) or the observation period, depending on \texttt{Status}. & Continuous & mean = 19.20, sd = 18.04, median = 12,
[Q1; Q3] = [6; 26], [min; max] = [1; 89].  \\ 
\texttt{Age} & Patient's age at hospital admission. &Continuous & mean = 81.36, sd = 8.27, median = 82.0, [Q1; Q3] = [76; 88], [min; max] = [60; 106] \\
\texttt{Gender}& Patient's biological sex (male/female). &Categorical & 55.83\% for male, 44.17\% for female.\\ 
\texttt{ModMCS} & Modified Multisource-Comorbidity Score \citep{caldera2025uncovering, corrao2017developing} quantifying cumulative burden of other diseases. &Continuous &mean = 11.47, sd = 7.77, median = 10, [Q1; Q3] = [6; 16], [min; max] = [0; 57] \\ 
\texttt{Resp} & Count of respiratory diseases among \textit{chronic obstructive pulmonary disease}, \textit{bronchitis}, \textit{pneumonia}, and \textit{respiratory failure}. 
& Continuous & mean = 0.97, sd = 1.18, median = 1, [Q1; Q3] = [0; 2], [min; max] = [0; 4] \\

\bottomrule
\end{tabularx}
\end{table}

\subsection{Model setup and results}

We fit our model under multiple configurations and identify the optimal one as that with the maximum mean silhouette value, as described in Section~\ref{sec:estimationdetails} \citep{kaufman2009finding}.
Specifically, we consider two choices for the frailty distribution (Gamma and Inverse Gaussian) and three options for the baseline hazard distribution (Weibull and Exponential).
For the penalization parameters, we explore $\gamma = \{10^{-8}, 10^{-5}, 10^{-2}, 10^{-1}\}$ and $k \in \{ 50, 100\}$.
Finally, we evaluate models with $C \in \{2, 3, 4\}$.
We set \texttt{maxit} = 200, \texttt{tolS} = $10^{-2}$, \texttt{tolll} = $1$ and leave all the other parameters with default settings.
The mean silhouette values for all the above configurations are reported in Table~\ref{tab:CaseStudySilhouette}.

\begin{table}[!htbp] 
\centering
\footnotesize
\begin{tabular}{cccccccc}
\toprule
\multicolumn{2}{c}{Parametric Distributions} & \multicolumn{2}{c}{Penalization Param.} & \multicolumn{3}{c}{$C$} \\
\cmidrule(lr){1-2} \cmidrule(lr){3-4} \cmidrule(lr){5-7}
$u_g$ & $H_0(t; \boldsymbol{\psi})$ & $\gamma$ & $k$ & 2 & 3 & 4 \\
\toprule
\multirow{8}{*}{Gamma} & \multirow{8}{*}{Weibull} 
   & \multirow{2}{*}{$10^{-8}$} &  50 & 0.544 & 0.470 & 0.320 \\
&  & & 150 &  0.535 & 0.526 & 0.277 \\
\cmidrule(lr){3-7}
 &  & \multirow{2}{*}{$10^{-5}$} &  50 & 0.542 & 0.379 & 0.416 \\
&  & & 150 & 0.526 & 0.496 & 0.293 \\
\cmidrule(lr){3-7}
 &  & \multirow{2}{*}{$10^{-2}$} &  50 & 0.526 & 0.508 & 0.292 \\
&  & & 150 & 0.518 & 0.464 & 0.461 \\
\cmidrule(lr){3-7}
 &  & \multirow{2}{*}{$10^{-1}$} &  50 & 0.474 & 0.363 & 0.252 \\
&  & & 150 & 0.591 & \textbf{0.608} & 0.033 \\
\midrule
\multirow{8}{*}{Gamma} & \multirow{8}{*}{Exponential} 
   & \multirow{2}{*}{$10^{-8}$} &  50 & 0.542 & 0.376 & 0.243 \\
&  & & 150 & 0.548 & 0.473 & 0.196 \\
\cmidrule(lr){3-7}
 &  & \multirow{2}{*}{$10^{-5}$} &  50 & 0.545 & 0.290 & 0.277 \\
&  & & 150 & 0.533 & 0.514 & 0.092 \\
\cmidrule(lr){3-7}
 &  & \multirow{2}{*}{$10^{-2}$} &  50 & 0.538 & 0.495 & 0.440\\
&  & & 150 & 0.525 & 0.404 & 0.379 \\
\cmidrule(lr){3-7}
 &  & \multirow{2}{*}{$10^{-1}$} &  50 &  0.524 & 0.126 & 0.441 \\
&  & & 150 &  0.493 & 0.132 & 0.402 \\
\midrule
\multirow{8}{*}{Inverse Gaussian} & \multirow{8}{*}{Weibull} 
   & \multirow{2}{*}{$10^{-8}$} &  50 & 0.545 & 0.512 & 0.399 \\
&  & & 150 & 0.545 & 0.498 & 0.419 \\
\cmidrule(lr){3-7}
 &  & \multirow{2}{*}{$10^{-5}$} &  50 & 0.525 & 0.462 & 0.170 \\
&  & & 150 &  0.527 & -0.244 & 0.187\\
\cmidrule(lr){3-7}
 &  & \multirow{2}{*}{$10^{-2}$} &  50 & 0.532 & 0.491 & 0.426 \\
&  & & 150 & 0.528 & 0.420 & 0.412 \\
\cmidrule(lr){3-7}
 &  & \multirow{2}{*}{$10^{-1}$} &  50 & 0.487 & 0.481 & 0.224 \\
&  & & 150 &  0.594 & \textbf{0.609} & 0.001 \\
\midrule
\multirow{8}{*}{Inverse Gaussian} & \multirow{8}{*}{Exponential} 
   & \multirow{2}{*}{$10^{-8}$} &  50 & 0.544 & 0.452 & 0.434 \\
&  & & 150 & 0.546 & 0.485 & 0.473 \\
\cmidrule(lr){3-7}
 &  & \multirow{2}{*}{$10^{-5}$} &  50 & 0.543 & 0.286 & 0.472 \\
&  & & 150 &  0.525 & 0.305 & 0.417 \\
\cmidrule(lr){3-7}
 &  & \multirow{2}{*}{$10^{-2}$} &  50 & -0.228 & 0.472 & 0.514 \\
&  & & 150 & 0.533 & 0.437 & 0.427 \\
\cmidrule(lr){3-7}
 &  & \multirow{2}{*}{$10^{-1}$} &  50 & 0.536 & 0.046 & 0.209 \\
&  & & 150 & 0.502 & 0.141 & 0.126 \\
\bottomrule
\end{tabular}
\caption{\small Mean silhouette values across model configurations, varying frailty and baseline hazard distributions, penalization parameters $\gamma$ and $k$, and number of clusters $C$.}
\label{tab:CaseStudySilhouette}
\end{table}

\medskip


We select the model with the frailty distributed as Inverse Gaussian, the baseline as Weibull, $\gamma = 10^{-1}$, $k = 150$ and $C=3$ as the optimal clustering configuration, as it yields the highest mean silhouette score equal to 0.609 and therefore the best separation with respect to the fitted survival model. 
With the chosen model we obtain the following regression coefficients: $\hat\beta_{\texttt{Age}}=0.0025$, $\hat\beta_{\texttt{Gender}=\text{male}}=0.0244$, $\hat\beta_{\texttt{ModMCS}}=0.0013$, and $\hat\beta_{\texttt{Resp}} = 0.0033$.
The corresponding p-values are all $<0.05$, indicating statistically significant effects.
The estimated frailty variance is $\theta = 0.573$, while the Weibull shape and scale parameters were $\rho = 0.618 $ and $\lambda = \psi^\rho = 1.126$, respectively.

The model reaches convergence in 23 iterations.
The three obtained clusters -- corresponding to disconnected components in the patients graph -- contain, respectively, 1077, 1054 and 955 patients, revealing subgroups with differing risk profiles.
Specifically, the first (and largest) cluster includes older patients (mean age 83.5 years) with the highest modified multisource-comorbidity score (average \texttt{ModMCS} equal to 12.6) while moderate respiratory disease burden (average \texttt{Resp} equal to 0.99), predominantly males, and has a mean follow-up time of 0.147.
Cluster 2 comprises slightly younger patients (mean age 81.8 years) with intermediate \texttt{ModMCS} (11.8) and the highest respiratory disease burden (1.03), also male-predominant, with a mean follow-up of 0.158.
Cluster 3, the smallest one, contains the youngest patients (mean age 78.4 years) with the lowest \texttt{ModMCS} (9.8) and respiratory disease burden (0.87), more females than males, and had the longest mean follow-up (0.176).

Cluster membership varied substantially across hospitals. The median proportion of observations belonging to the dominant cluster was 75.2\% (interquartile range 66.9–95.5\%), indicating moderate within-hospital heterogeneity overall. Thirteen hospitals (40.6\%) showed a strong predominance of a single cluster, with at least 80\% of observations assigned to that cluster, while five hospitals (15.6\%) were entirely associated with one cluster. In contrast, only three hospitals (9.4\%) exhibited a mixed cluster composition, with no cluster accounting for more than 60\% of observations.

The wide range in cluster patterns -- from hospitals dominated by a single cluster to those with a more mixed composition -- underscores substantial within-hospital heterogeneity and supports using an approach specifically designed to handle this variability rather than assuming uniformity. This heterogeneity is further reflected in the estimated $\theta$, indicating meaningful variation in hospital-level effects that our method accounts for.

\section{Discussion}
\label{sec:discussion}

Motivated by the need to identify latent patient subgroups in hierarchical time-to-event data, in this work we propose a framework that jointly performs patient risk-based clustering and survival modeling while accounting for group-level dependence through a parametric shared frailty model, ensuring that cluster assignments are informed by underlying survival dynamics rather than by covariate similarity alone.
From a methodological perspective, our approach addresses a gap in the existing literature by jointly modeling survival outcomes, latent patient subgroups, and group-level dependence within a single likelihood-based framework, allowing patient stratification to emerge from estimated risk profiles while properly accounting for within-hospital correlation.

The simulation study demonstrates that the proposed methodology reliably recovers latent cluster structures across a broad range of penalization settings, provided that the frailty variance is kept small so that cluster-specific effects remain the dominant source of heterogeneity. This choice facilitates identifiability by preventing individual-level random effects from obscuring systematic differences in underlying risk trajectories. Under these conditions, the method also achieves accurate estimation of the parametric frailty model when penalization is appropriately tuned using the silhouette coefficient as the selection criterion for both optimal clustering and parameter estimation.

When applied to the Enhance-Heart cohort of heart-failure patients hospitalized for COVID-19, the method identifies three  clusters of patients with distinct risk profiles and clinically interpretable differences in age, comorbidity burden, respiratory conditions, and survival patterns. The clusters were obtained by selecting the parameter configuration and parametric distributions that maximize the silhouette coefficient, ensuring both optimal separation and cohesion. These clusters correspond to disconnected components in the patient graph and reveal clinically meaningful subgroups: the first cluster captures older patients with higher comorbidity but moderate respiratory burden, the second includes slightly younger patients with the highest respiratory burden, and the third comprises the youngest patients with the lowest comorbidity and respiratory burden, highlighting how the methodology can uncover nuanced heterogeneity in event risk patterns.

Beyond patient-level stratification, our results highlight substantial heterogeneity across hospitals. Cluster composition varied markedly between hospitals, with many institutions showing a strong predominance of a single cluster, while a smaller subset exhibited more heterogeneous patient mixes. Furthermore, the non-negligible value of the frailty parameter suggests that latent institutional factors -- such as resource availability, clinical protocols, or contextual pressures during the COVID-19 pandemic -- may influence survival outcomes beyond what is captured by observed covariates and cluster membership. This finding underscores the importance of modeling hierarchical dependence explicitly and supports the combined use of clustering and frailty terms as complementary tools for disentangling multiple sources of heterogeneity.

\medskip

Some limitations warrant consideration. 
First, the construction of the similarity graph relies on the choice of the neighborhood size parameter $k$, which influences graph sparsity and cluster formation. Although adaptive tuning of the penalization parameter $\lambda$ embedded in the algorithm mitigates this issue in practice, the relationship between 
$k$ and the target number of clusters $C$ remains an area for further investigation. 
Furthermore, each unit is forced to connect to its $k$ nearest neighbors regardless of whether these lie near the cluster core or at its boundary. This limitation has been noted in the spectral clustering literature \citep{cai2022new}, where pruning weak connections and modifying the optimization term through power-function asymptotics have been proposed as potential remedies. However, such approaches introduce additional tuning parameters, including a maximum neighborhood size, and did not yield performance improvements in our setting.

In addition, in the current framework, a common baseline hazard is assumed across clusters, which is simplified in the penalization distance, while heterogeneity is captured through covariates and frailty terms. Future work will aim to extend the model to allow cluster-specific baseline hazards, enabling a more nuanced characterization of risk dynamics across patient subgroups. This extension would require reconsidering the frailty structure, as different clusters would have distinct frailties, going beyond the scopes of the present work. Additional developments could explore alternative similarity measures, as well as extensions to semi-parametric or fully nonparametric baseline hazards, further increasing the model's flexibility.

\subsection*{Competing interests}
No competing interest is declared.

\subsection*{Data availability}
The participants of this study did not give written consent for their data to be shared publicly, so due to the sensitive nature of the research and regulations applied from RL sharing terms, the full supporting data is not available.

\subsection*{Acknowledgements}
This work is part of the \textit{Enhance-Heart} project: Efficacy evaluatioN of the therapeutic-care patHways, of the
heAlthcare providers effects aNd of the risk stratifiCation in patiEnts suffering from HEART failure. The authors acknowledge
the ‘Unità Organizzativa Osservatorio Epidemiologico Regionale’ and ARIA S.p.A for providing data and technological support, and the financial support from the Department of Mathematics of Politecnico di Milano through the grant \lq Dipartimento di Eccellenza 2023-2027\rq \, by MUR, Italy.

\begin{appendix}
\counterwithin{equation}{section}
\counterwithin{figure}{section}
\counterwithin{table}{section}
\renewcommand\theequation{\thesection\arabic{equation}}

\section{Derivation of the marginal likelihood}
\label{app:prooflikel}

The \textit{conditional likelihood} for the $g$-th cluster is expressed as 
\begin{equation*}
    L_g^c(\boldsymbol{\psi}, \boldsymbol{\beta} \mid u_g, \mathcal{Z}) = \prod_{i=1}^{n_g} \left( h_0(y_{gi}) u_g \exp(\boldsymbol{x}_{gi}' \boldsymbol{\beta}) \right)^{\delta_{gi}} 
\cdot \exp\left( - H_0(y_{gi}) u_g \exp(\boldsymbol{x}_{gi}' \boldsymbol{\beta}) \right) 
\end{equation*}
and the \textit{marginal likelihood} is obtained by integrating out the frailty distribution over its support $S_U$:
\begin{align*}
    L_{g}(\boldsymbol{\psi}, \boldsymbol{\beta}, \theta, \mathcal{Z}) & = \int_{S_U} L_g^c(\boldsymbol{\psi}, \boldsymbol{\beta} \mid u_g, \mathcal{Z}) \cdot f_U(u_g \mid \theta) du_g \\
    & = \int_{S_U} \prod_{i=1}^{n_g} \left( h_0(y_{gi}) u_g \exp(\boldsymbol{x}_{gi}' \boldsymbol{\beta}) \right)^{\delta_{gi}} 
\cdot \exp\left( - H_0(y_{gi}) u_g \exp(\boldsymbol{x}_{gi}' \boldsymbol{\beta}) \right)  \cdot f_U(u_g \mid \theta) du_g \\ 
& = \prod_{i=1}^{n_g} \left( h_0(y_{gi}) \exp(\boldsymbol{x}_{gi}' \boldsymbol{\beta}) \right)^{\delta_{gi}} \int_{S_U} \prod_{i=1}^{n_g} \left( u_g \right)^{\delta_{gi}} 
\cdot \exp\left( - H_0(y_{gi}) u_g \exp(\boldsymbol{x}_{gi}' \boldsymbol{\beta}) \right)  \cdot f_U(u_g) du_g \\
& = \prod_{i=1}^{n_g} \left( h_0(y_{gi}) \exp(\boldsymbol{x}_{gi}' \boldsymbol{\beta}) \right)^{\delta_{gi}} \int_{S_U}  u_g ^{\sum_{i=1}^{n_g} \delta_{gi}} 
\cdot \exp\left( - \sum_{i=1}^{n_g} H_0(y_{gi}) \exp(\boldsymbol{x}_{gi}' \boldsymbol{\beta})  u_g \right)  \cdot f_U(u_g) du_g \\
& = \prod_{i=1}^{n_g} \left( h_0(y_{gi}) \exp(\boldsymbol{x}_{gi}' \boldsymbol{\beta}) \right)^{\delta_{gi}} (-1)^{d_g} \mathcal{L}^{(d_g)}\bigg( \sum_{i=1}^{n_g} H_0(y_{gi}) \exp(\boldsymbol{x}_{gi}' \boldsymbol{\beta}) \bigg)
\end{align*}
where in the last equation we used $(-1)^k\mathcal{L}^{(k)}(s) := (-1)^k \frac{d^k}{ds^k}\mathcal{L}(s) : = \mathbb{E}[U^k\exp(-Us)]$, being  $\mathcal{L}(s)$, for $s\geq0$, the Laplace transform \citep{munda2014adjusting, klein1992semiparametric}.
By taking the logarithm and summing over the clusters, we obtain Eq. (\ref{eq:marginalloglik}).

\section{Derivation of closed-form solution of (\ref{eq:step3longform})}
\label{app:proofKKT}

Given that 
$\sum_{h,j} \mu s_{gi,hj}^2 + w_{gi,hj}s_{gi,hj} = \mu \| \mathbf{s}_{gi} \|^2 + \boldsymbol{w}_{gi}' \mathbf{s}_{gi} = \mu \| \mathbf{s}_{gi} + \frac{1}{2\mu}\boldsymbol{w}_{gi}\|^2 - \frac{1}{4\mu}\| \boldsymbol{w}_{gi}\|^2$, problem (\ref{eq:step3longform}) can be rewritten as
\begin{align*}
    \min_{\mathbf{s}_{gi}' \mathbf{1} = 1, \mathbf{s}_{gi} \geq 0} \quad 
& \big \| \mathbf{s}_{gi} + \frac{1}{2\mu} \boldsymbol{w}_{gi} \big\|^2
\end{align*}
where the last term $\frac{1}{4\mu}\| \boldsymbol{w}_{gi}\|^2$ and the multiplicative $\mu$ were removed as the former does not depend on $\mathbf{s}_{gi}$ and the latter can be removed from the optimization problem.
As we said that the problem is solved independently for each unit $i$ in group $g$, in this section from now on we denote $\mu$ as $\mu_{gi}$.
This problem can be solved through the KKT conditions.
Indeed, the Lagrangian function can be written as
$$ \mathcal{L} = \big\| \mathbf{s}_{gi} + \frac{1}{2\mu_{gi}} \boldsymbol{w}_{gi} \big\|^2 - \alpha_{gi} (\mathbf{s}_{gi}' \mathbf{1} - 1) - \boldsymbol{r}' \mathbf{s}_{gi} $$
where $\alpha_{gi} \geq 0$ and $\boldsymbol{r} \geq 0$ are Lagrangian multipliers.
According to the KKT condition, we have the following solution 
\begin{equation}
    s_{gi,hj} = \max\bigg\{0, \alpha_{gi} - \frac{w_{gi,hj}}{2\mu_{gi}} \bigg\}
    \label{eq:generalsolutionforS}
\end{equation}
subject to $\mathbf{s}_{gi}' \mathbf{1} = 1$.

\section{Derivation of \texorpdfstring{$\alpha_{gi}$}{alpha gi} and \texorpdfstring{$\mu_{gi}$}{mu gi} in (\ref{eq:alpha_mu})}
\label{app:getalphamu}

The optimal solution is given by \ref{eq:generalsolutionforS},
subject to the constraint $\mathbf{s}_{gi}' \mathbf{1} = \sum_{h=1}^N s_{gi,hj}  = 1$.

Let $\hat w_{gi,1} \le \hat w_{gi,2} \le \cdots \le \hat w_{gi,N-1}$ denote the sorted distances
(excluding the self-distance).
Imposing $k$-nearest-neighbor sparsity yields
\begin{align*}
\hat s_{gi,p} &> 0, \quad p = 1,\ldots,k, \\
\hat s_{gi,p} &= 0, \quad p = k+1,\ldots,N .
\end{align*}

For the $k$ nonzero entries,
\begin{equation*}
\hat s_{gi,p}
=
\alpha_{gi} - \frac{\hat w_{gi,p}}{2\mu_{gi}},
\qquad p = 1,\ldots,k.
\end{equation*}

For the $(k+1)$-th neighbor, the sparsity condition implies
\begin{equation*}
\alpha_{gi} - \frac{\hat w_{gi,k+1}}{2\mu_{gi}} \le 0.
\end{equation*}
At optimality this constraint is active, giving
\begin{equation*}
\label{eq:alpha_threshold}
\alpha_{gi} = \frac{\hat w_{gi,k+1}}{2\mu_{gi}} .
\end{equation*}

Enforcing the simplex constraint over the $k$ nonzero entries yields
\begin{align*}
\sum_{p=1}^k \hat s_{gi,p}
&=
\sum_{p=1}^k
\left(
\alpha_{gi} - \frac{\hat w_{gi,p}}{2\mu_{gi}}
\right)
= 1,
\\
k\alpha_{gi}
-
\frac{1}{2\mu_{gi}}
\sum_{p=1}^k \hat w_{gi,p}
&= 1.
\end{align*}
Solving for $\alpha_{gi}$ gives
\begin{equation}
\label{eq:alpha}
\alpha_{gi}
=
\frac{1}{k}
+
\frac{1}{2k\mu_{gi}}
\sum_{p=1}^k \hat w_{gi,p}.
\end{equation}

Combining \eqref{eq:alpha_threshold} and \eqref{eq:alpha}, we obtain
\begin{align*}
\frac{\hat w_{gi,k+1}}{2\mu_{gi}}
&=
\frac{1}{k}
+
\frac{1}{2k\mu_{gi}}
\sum_{p=1}^k \hat w_{gi,p},
\\
2\mu_{gi}
&=
k\hat w_{gi,k+1}
-
\sum_{p=1}^k \hat w_{gi,p}.
\end{align*}
Therefore,
\begin{equation*}
\mu_{gi}
=
\frac{k}{2}\hat w_{gi,k+1}
-
\frac{1}{2}
\sum_{p=1}^k \hat w_{gi,p}.
\end{equation*}


\section{Parameter estimates: variability and estimation error}
\label{app:msevariance}

In this Appendix, we interpret Table~\ref{tab:MSE_Variance}.
The table reports the normalized mean squared error and variance of the parameter estimates for different penalization levels $\gamma$ under both administrative and normal censoring. The estimators remain stable for very small penalties ($\gamma \leq 10^{-4}$), with MSE/Var values close to one across all parameters, indicating that estimation error is almost entirely driven by variance—an observation also confirmed by Figures~\ref{fig:all_estimates_admin} and \ref{fig:all_estimates_norm}. As $\gamma$ increases into the range $10^{-3}$–$10^{-2}$, bias begins to emerge, most noticeably for $\rho$ and $\beta$, where MSE/Var rises despite relatively modest changes in variance. For larger penalties ($\gamma \geq 0.1$), estimation accuracy deteriorates sharply, with substantial inflation of MSE/Var, particularly for $\rho$ and $\beta$. This pattern is observed under both censoring mechanisms but is consistently more pronounced when censoring occurs according to a normal distribution.

These findings should be interpreted jointly with Table~\ref{tab:AccuracyandARI}, where an intermediate penalization level emerges as optimal, reflecting a trade-off between shrinkage and estimation error.

\begin{table}[!htbp] 
\centering
\footnotesize
\begin{tabular}{ll cc cc cc cc}
\toprule
\textbf{Censoring} & $\gamma$ & \multicolumn{2}{c}{$\theta$} & \multicolumn{2}{c}{$\rho$} & \multicolumn{2}{c}{$\psi$}  & \multicolumn{2}{c}{$\beta$}  \\
\cmidrule(lr){3-4} \cmidrule(lr){5-6} \cmidrule(lr){7-8} \cmidrule(lr){9-10}
& & MSE/Var & Var &  MSE/Var & Var &  MSE/Var & Var &  MSE/Var & Var \\
\toprule
\multirow{8}{*}{(i) Administrative} 
  &  $0$ & 1.022 & 0.051 & 1.006 & 0.013 & 1.001 & 0.000 & 1.007 & 0.001 \\
  &  $10^{-6}$ & 1.022 & 0.051 & 1.006 & 0.013 & 1.001 & 0.000 & 1.007 & 0.001 \\
  &  $10^{-4}$ & 1.023 & 0.051 & 1.008 & 0.013 & 1.001 & 0.000 & 1.010 & 0.001 \\
  &  $10^{-3}$ & 1.025 & 0.050 & 1.042 & 0.012 & 1.003 & 0.000 & 1.056 & 0.001 \\
  &  $10^{-2}$ & 1.046 & 0.046 & 2.633 & 0.011 & 1.077 & 0.000 & 3.190 & 0.001 \\
  &  $0.1$ & 1.000 & 0.069 & 20.407 & 0.025 & 3.036 & 0.000 & 20.207 & 0.003 \\
  &  $0.2$ & 1.047 & 0.091 & 35.186 & 0.027 & 4.269 & 0.000 & 33.404 & 0.003 \\
  &  $0.4$ & 1.205 & 0.096 & 44.550 & 0.034 & 5.471 & 0.000 & 43.941 & 0.004 \\

\midrule
\multirow{8}{*}{(ii) Normal} 
  &  $0$ &  1.022 & 0.052 & 1.004 & 0.012 & 1.002 & 0.000 & 1.007 & 0.001 \\
  &  $10^{-6}$ & 1.022 & 0.052 & 1.004 & 0.012 & 1.002 & 0.000 & 1.007 & 0.001 \\
  &  $10^{-4}$ & 1.024 & 0.052 & 1.009 & 0.012 & 1.003 & 0.000 & 1.013 & 0.001 \\
  &  $10^{-3}$ & 1.036 & 0.050 & 1.131 & 0.011 & 1.007 & 0.000 & 1.169 & 0.001 \\
  &  $10^{-2}$ & 1.170 & 0.041 & 5.808 & 0.010 & 1.137 & 0.000 & 6.310 & 0.001 \\
  &  $0.1$ & 1.077 & 0.054 & 46.782 & 0.014 & 3.922 & 0.000 & 39.592 & 0.002 \\
  &  $0.2$ & 1.001 & 0.068 & 77.124 & 0.014 & 5.371 & 0.000 & 64.944 & 0.002 \\
  &  $0.4$ & 1.058 & 0.078 & 74.959 & 0.022 & 6.449 & 0.000 & 71.004 & 0.003 \\

\bottomrule
\end{tabular}
\caption{\small 
Normalized mean squared error (MSE/Var) and variance (Var) of the parameter estimates for $k=20$ and $C=3$.
For each parameter $p \in {\theta, \rho, \xi, \beta}$, we define
$\text{MSE}(\hat{p}) = \mathbb{E}[\|\hat{p} - p\|_2^2]$ and
$\text{Var}(\hat{p}) = \mathbb{E}[\|\hat{p} - \mathbb{E}[\hat{p}]\|_2^2]$,
with $p$ denoting the true parameter introduced in Section~\ref{sec:DGP}.
The table reports results under administrative and normal censoring for all considered penalization levels $\gamma$.}

\label{tab:MSE_Variance}
\end{table}

\end{appendix}

\bibliographystyle{chicago} 
\bibliography{bibliography}

\end{document}